\documentclass[10pt,journal,twocolumn,twoside]{IEEEtran}
\usepackage{epsfig,amsfonts,subfigure,color,amsbsy}
\usepackage[nolist]{acronym}
\usepackage{graphicx,cite,amssymb,amsthm,bm}
\usepackage{amsmath}
\usepackage{mathrsfs}
\usepackage{bm,bbm}
\usepackage{multirow}
\usepackage{bigstrut}
\usepackage{psfrag}
\usepackage{stackengine}
\usepackage{xcolor}
\usepackage{mathtools}
\usepackage{tabularx,array}
\usepackage{mathbbol}
\usepackage{lipsum}
\usepackage[sans]{dsfont}
\mathtoolsset{showonlyrefs}
\usepackage{array, makecell}
\usepackage{float}
\usepackage{balance}
\usepackage{verbatim}
\usepackage{mathrsfs}
\usepackage{tikz}
\usepackage{pgfplots}
\usepackage{scalerel}

\pgfplotsset{compat=1.17}

\usepackage{tikz}
\usetikzlibrary{shapes,arrows}
\usetikzlibrary{calc}
\newtheorem{example}{Example}

\definecolor{TUMBlack}{cmyk}{0,0,0,1}     
\definecolor{TUMWhite}{cmyk}{0,0,0,0}     

\definecolor{TUMBlue} {cmyk}{1,0.43,0,0}  

\definecolor{TUMDarkBlue}   {cmyk}{1,0.57,0.12,0.7}      
\definecolor{TUMDarkerBlue} {cmyk}{1,0.54,0.04,0.19}     
\definecolor{TUMMediumBlue} {cmyk}{0.9,0.48,0,0}         
\definecolor{TUMLighterBlue}{cmyk}{0.65,0.19,0.01,0.04}  
\definecolor{TUMLightBlue}  {cmyk}{0.42,0.09,0,0}        

\definecolor{TUMDarkGray}  {cmyk}{0,0,0,0.8}  
\definecolor{TUMMediumGray}{cmyk}{0,0,0,0.5}  
\definecolor{TUMLightGray} {cmyk}{0,0,0,0.2}  

\definecolor{TUMGreen} {cmyk}{0.35,0,1,0.2}         
\definecolor{TUMOrange}{cmyk}{0,0.65,0.95,0}        
\definecolor{TUMIvory} {cmyk}{0.03,0.04,0.14,0.08}  

\definecolor{TUMBeamerYellow}    {rgb}{1.00,0.71,0.00}  
\definecolor{TUMBeamerOrange}    {rgb}{1.00,0.50,0.00}  
\definecolor{TUMBeamerRed}       {rgb}{0.90,0.20,0.09}  
\definecolor{TUMBeamerDarkRed}   {rgb}{0.79,0.13,0.25}  
\definecolor{TUMBeamerBlue}      {rgb}{0.00,0.60,1.00}  
\definecolor{TUMBeamerLightBlue} {rgb}{0.25,0.75,1.00}  
\definecolor{TUMBeamerGreen}     {rgb}{0.57,0.67,0.42}  
\definecolor{TUMBeamerLightGreen}{rgb}{0.71,0.79,0.51}  

\DeclareMathOperator*{\argmax}{arg\,max}

\newcommand{\rmin}{\underline{R}}
\newcommand{\rmax}{\overline{R}}
\newcommand{\omin}{\underline{\omega}}
\newcommand{\omax}{\overline{\omega}}
\newcommand{\rset}{\mathcal{R}}

\newcommand{\thr}{\gamma^\star}

\newcommand{\inR}{R_{\scaleto{\mathsf{I}}{3.5pt}}}
\newcommand{\inmR}{R_{\scaleto{\mathsf{IM}}{3.5pt}}}
\newcommand{\dmR}{R_{\scaleto{\mathsf{O}}{3.5pt}}}
\newcommand{\Hb}{\mathsf{H}_{\scaleto{\mathsf{b}}{4pt}}}
\newcommand{\incode}{\mathcal{C}_{\scaleto{\mathsf{I}}{3.5pt}}}
\newcommand{\inmcode}{\mathcal{C}_{\scaleto{\mathsf{IM}}{3.5pt}}}
\newcommand{\outcode}{\mathcal{C}_{\scaleto{\mathsf{O}}{3.5pt}}}
\newcommand{\code}{\mathcal{C}}
\newcommand{\PEP}{\mathsf{PEP}}
\newcommand{\prob}[1]{\mathsf{P}\!\left[#1\right]}
\newcommand{\expect}[1]{\mathsf{E}\!\left[#1\right]}
\newcommand{\iowe}{A^{\scaleto{\mathsf{IO}}{3.5pt}}}
\newcommand{\fieldtwo}{\mathbbmss{F}_{\!\scaleto{\mathsf{2}}{3.5pt}}}
\newcommand{\liftfac}{\ell}
\newcommand{\hw}{w_{\scaleto{\mathsf{H}}{3.5pt}}}
\newcommand{\hd}{d_{\scaleto{\mathsf{H}}{3.5pt}}}

\newcommand{\cbsc}{C_{\scaleto{\mathsf{BSC}}{3.5pt}}}
\newcommand{\cawgn}{C_{\scaleto{\mathsf{AWGN}}{3.5pt}}}

\newcommand{\vecv}{\bm{v}}
\newcommand{\vecx}{\bm{x}}
\newcommand{\vecy}{\bm{y}}
\newcommand{\vecw}{\bm{w}}
\newcommand{\vecc}{\bm{c}}
\newcommand{\vecz}{\bm{z}}

\newcommand{\vecb}{\bm{b}}

\newcommand{\vecZ}{\bm{Z}}
\newcommand{\vecY}{\bm{Y}}

\newcommand{\ensP}[2]{\mathscr{C}_{#2}\!\left(#1\right)}
\newcommand{\ensfP}[1]{\mathscr{C}\!\left(#1\right)}
\newcommand{\pgraph}{\mathcal{P}}
\newcommand{\bgraph}{\mathcal{G}}

\newcommand{\vn}{\mathsf{v}}
\newcommand{\cn}{\mathsf{c}}
\newcommand{\transposed}{{\scaleto{\mathsf{T}}{3.5pt}}}

\newcommand{\vB}{\bm{B}}
\newcommand{\vH}{\bm{H}}
\newcommand{\vX}{\bm{X}}

\newcommand{\xML}{\hat{\vecx}_{\scaleto{\mathsf{ML}}{3.5pt}}}
\newcommand{\xMM}{\hat{\vecx}_{\scaleto{\mathsf{MM}}{3.5pt}}}

\IEEEoverridecommandlockouts
\begin{document}

\makeatletter
\newcommand*{\encircled}[1]{\relax\ifmmode\mathpalette\@encircled@math{#1}\else\@encircled{#1}\fi}
\newcommand*{\@encircled@math}[2]{\@encircled{$\m@th#1#2$}}
\newcommand*{\@encircled}[1]{%
  \tikz[baseline,anchor=base]{\node[draw,circle,outer sep=0pt,inner sep=.2ex] {#1};}}
\makeatother

\title{Rate-Adaptive Protograph MacKay-Neal Codes}

\author{
Ayman Zahr, Balazs Matuz, Gianluigi Liva
\thanks{
		A. Zahr, B. Matuz, and G. Liva are  with the Institute of Communications and Navigation, German Aerospace Center (DLR), Wessling, Germany (email:  \{ayman.zahr,balazs.matuz,gianluigi.liva\}@dlr.de).
}
\thanks{The authors acknowledge the financial support by the Federal Ministry of Education and Research of Germany in the programme of "Souver\"an. Digital. Vernetzt." Joint project 6G-RIC, project identification number: 16KISK022.
}
}

 \maketitle
 \thispagestyle{empty}
\pagestyle{empty}
 \IEEEoverridecommandlockouts
 


\begin{abstract}
A class of rate-adaptive protograph MacKay-Neal (MN) codes is introduced and analyzed. The code construction employs an outer distribution matcher (DM) to adapt the rate of the scheme. The DM is coupled with an inner protograph-based low-density parity-check (LDPC) code, whose base matrix is optimized via density evolution analysis to approach the Shannon limit of the binary-input additive white Gaussian noise (biAWGN) channel over a given range of code rates. The density evolution analysis is complemented by finite-length simulations, and by a study of the  error floor performance.
\end{abstract}



\pagenumbering{arabic}


\begin{acronym}
    \acro{BEC}{binary erasure channel}
    \acro{BP}{belief propagation}
    \acro{DE}{density evolution}
    \acro{LDPC}{low-density parity-check}
    \acro{ML}{maximum likelihood}
    \acro{r.v.}{random variable}
    \acro{PEP}{pairwise error probability}
    \acro{BP}{belief propagation}
    \acro{BPSK}{binary phase shift keying}
    \acro{BSC}{binary symmetric channel}
    \acro{AWGN}{additive white Gaussian noise}
    \acro{OOK}{on-off keying}
    \acro{DM}{distribution matcher}
    \acro{p.m.f.}{probability mass function}
    \acro{p.d.f.}{probability density function}
    \acro{i.i.d.}{independent and identically-distributed}
    \acro{CC}{constant composition}
    \acro{LEO}{low earth orbit}
    \acro{biAWGN}{binary-input additive white Gaussian noise}
    \acro{PAM}{pulse amplitude modulation}
    \acro{SNR}{signal-to-noise ratio}
    \acro{EXIT}{extrinsic information transfer}
    \acro{PEXIT}{protograph extrinsic information transfer}
    \acro{VN}{variable node}
    \acro{CN}{check node}
    \acro{FER}{frame error rate}
    \acro{MN}{MacKay-Neal}
    \acro{RA}{repeat-accumulate}
    \acro{NS}{non-systematic}
    \acro{LLR}{log-likelihood ratio}
    \acro{MM}{mismatched}
    \acro{EPC}{\emph{equivalent parallel channel}}
    \acro{WCL}{worst-case loss}
    \acro{UB}{union bound}
    \acro{TUB}{truncated union bound}
    \acro{PEG}{progressive edge growth}
    \acro{PAS}{probabilistic amplitude shaping}
    \acro{CCDM}{constant composition distribution matcher}
    \acro{MI}{mutual information}
\end{acronym}

\section{Introduction}\label{sec:intro}

\ac{MN} codes  were introduced in \cite[Sec. VI]{Mac99} as a class of error correcting codes based on sparse matrices for nonuniform sources. 
They can be seen as multi-edge-type  \ac{LDPC} codes \cite{Gal63,RU08} defined by an extended bipartite graph with a set of \acp{VN} associated with the source bits, and the remaining \acp{VN} associated with the codeword bits. \ac{LDPC} code constructions closely related to \ac{MN} codes were proposed in \cite{Wainwright,fresia2010joint} for joint source and channel coding. While originally introduced to deal with nonuniform sources, it was pointed out in \cite{Mac99} that \ac{MN} can be employed also with uniform sources, by introducing a nonlinear block code that turns the uniform source output sequence into a sequence with a prescribed distribution. In   \cite{Mac99} it was also recognized that this option may be appealing from a rate flexibility viewpoint, since the rate of the scheme may be modified (e.g., adapted to varying channel conditions) by changing the statistics of the sequences produced by the nonlinear block encoder, hence without modifying the underlying \ac{LDPC} code. This approach is interesting from two viewpoints. First, the possibility of adapting the rate without changing the underlying \ac{LDPC} encoder/decoder is important to limit the implementation complexity (modern communication standards that seek rate flexibility often specify a large number of \ac{LDPC} code parity-check matrices to serve different rates, see e.g. \cite{Morello}). Second, a rate-adaptive scheme based on \ac{MN} codes allows to keep the block length constant, allowing the introduction of periodic synchronization markers, with benefits in terms of frame synchronization performance. Despite these appealing properties, \ac{MN} codes as a means to achieve rate flexibility received little attention. A notable exception is the \ac{PAS} scheme introduced in \cite{bocherer2015bandwidth}, where a construction reminiscent of \ac{MN} codes was proposed. In \ac{PAS}, the sequence output by a uniform (binary) source is also processed by the nonlinear block encoder, referred to as \ac{DM} \cite{bocherer2015bandwidth}, generating a sequence of amplitude symbols with a given empirical distribution. The binary labels associated with the amplitude symbols are encoded through a nonsystematic \ac{LDPC} code encoder, producing a parity bit vector. The amplitude symbols together with the parity bits are then mapped onto \ac{PAM} symbols. While the main result achieved by \ac{PAS} is to provide sizable shaping gains, it was quickly recognized that, as a byproduct, \ac{PAS} is naturally rate-adapting thanks to the possibility of tuning the \ac{DM} rate \cite{bocherer2015bandwidth}, as originally hypothesized in \cite{Mac99}. This aspect makes \ac{PAS} attractive for optical long haul transmission systems \cite{Buchali2016}. 

In this paper, we analyze and design a class of rate-adaptive \ac{MN} codes for the \ac{biAWGN} channel. The class is based on protograph \ac{LDPC} codes \cite{Tho03} coupled with a \ac{CCDM} \cite{bocherer2015bandwidth,Schulte2016}. Noting that the concatenation of the \ac{CCDM} with the inner linear block code results in a nonlinear code, we introduce an equivalent decoding problem. The equivalent problem resorts to the study of the performance of the protograph \ac{LDPC} code over the communication channel, where side information is provided to the decoder by observing the \ac{LDPC} encoder input through a binary-input, binary-output channel. By means of \ac{PEXIT} analysis \cite{ten01,Liva2007:EXIT_GC}, we design rate-adaptive \ac{MN} codes that are capable of operating close to the \ac{biAWGN} channel capacity over a wide range of code rates. A quantized \ac{DE} analysis \cite{RU01a} is used to confirm the accuracy of the thresholds computed via \ac{PEXIT} analysis. Finite-length performance results are finally included, and complemented by an error floor analysis that (for the designed codes) yields precise estimates of the code performance at large \acp{SNR}.

\begin{figure*}
	\begin{center}
		\includegraphics[width=16cm]{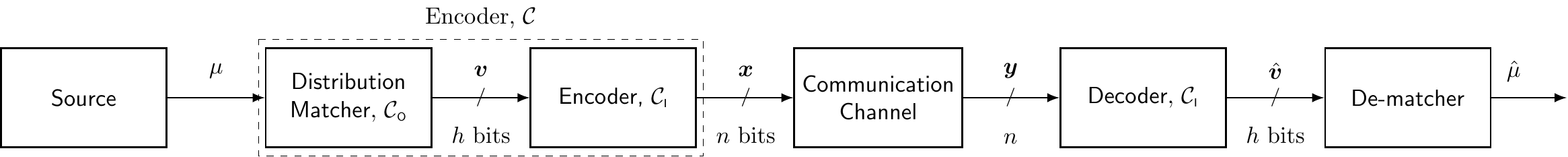}
        \vspace{-2mm}
		\caption{System model, where a \ac{MN} code is used to communicate over the \ac{biAWGN} channel (communication channel).}\label{fig:model}
	\end{center}
 \vspace{-3mm}
\end{figure*}

\section{Preliminaries}\label{sec:prel}

In the following, we denote \acp{r.v.} by uppercase letters and their realisations by lowercase letters. We denote the \ac{p.m.f.} of a discrete \ac{r.v.} $X$ as $P_X(x)= \prob{X=x}$, and the \ac{p.d.f.} of discrete \ac{r.v.} $X$ as $p_X(x)$. In either case, the subscript will be dropped whenever the context allows it, i.e., $P(x)=P_X(x)$ and $p(x)=p_X(x)$. We use $\Hb(\omega)$ to denote the binary entropy function, i.e., $\Hb(\omega)=-\omega \log_2 \omega -(1-\omega)\log_2(1-\omega)$ for $0<\omega<1$ and $\Hb(0)=\Hb(1)=0$. We refer to vectors as row vectors denoted by bold letters, e.g., $\bm{x}$, whereas matrices are denoted by uppercase bold letters, e.g., $\vX$. We denote by $\fieldtwo$ the order-$2$ finite field. We use $\hw(\bm{x})$ and $\hd(\bm{x}, \bm{y})$ to be respectively the Hamming weight of a vector $\bm{x}$ and the Hamming distance between two vectors $\bm{x}$ and $\bm{y}$. 

We consider transmission over the \ac{biAWGN} channel defined by $Y = X + N$ where $Y$ is the channel output, $X\in\{-1,+1\}$ is the channel input, and where $N\sim\mathcal{N}(0,\sigma^2)$ is the additive white Gaussian noise term. The channel \ac{SNR} is defined as $E_s/N_0 = 1/(2\sigma^2)$ where $E_s$ is the energy per symbol and $N_0$ is the single-sided noise power spectral  density.

\subsection{Protograph LDPC codes}\label{sec:prel:proto}

A protograph $\pgraph$ is a small bipartite graph consisting of a set of  $n_0$ \acp{VN}, a set of $m_0$ \acp{CN}, and $e$ edges. \acp{VN} in the protograph are numbered from $1$ to $n_0$. Similarly, protograph \acp{CN} are numbered from $1$ to $m_0$. Each \ac{VN}/\ac{CN}/edge in a protograph defines a \ac{VN}/\ac{CN}/edge type.   The bipartite graph $\bgraph$ of an \ac{LDPC} code can be derived by lifting the protograph. In particular, the protograph is copied ${\liftfac}$ times (where ${\liftfac}$ is referred to as the lifting factor), and the edges of the protograph copies are permuted under the following constraint: if an edge connects a type-$j$ \ac{VN} to a type-$i$ \ac{CN} in $\pgraph$, after permutation the edge should connect one of the ${\liftfac}$ type-$j$ \ac{VN} copies with one of the ${\liftfac}$ type-$i$ \ac{CN} copies in $\bgraph$. We denote by $\{\vn_1,\vn_2,\ldots,\vn_{n}\}$ the set of $n={\liftfac}n_0$ \acp{VN} in $\bgraph$, and by $\{\cn_1,\cn_2,\ldots,\cn_{m}\}$ the set of $m={\liftfac}m_0$ \acp{CN}.
The lifted graph $\bgraph$ defines the $m \times n$ parity-check matrix $\vH$.  The base matrix of a protograph is an $m_0 \times n_0$ matrix $\vB= [b_{i,j}]$ where $b_{i,j}$ is the number of edges that connect \ac{VN} $j$ to \ac{CN} $i$ in $\pgraph$. We will make use of \ac{LDPC} codes with \emph{punctured} (or \emph{state}) \acp{VN}. A punctured \ac{VN} is associated with a codeword bit that is not transmitted through the communication channel. We will assume that all the \acp{VN} of a given type are either punctured, or they are not. 

\section{Protograph MacKay-Neal Codes}\label{sec:MNcodes}

A \ac{MN} code can be constructed by concatenating an outer nonlinear code $\outcode$ with an inner code $\incode$ \cite{Mac99}. More specifically, we consider the setting depicted in Figure \ref{fig:model}. Here, an \ac{i.i.d.} uniform source generates a message $\mu \in \{1,2,\ldots,M\}$. The message is input to the encoder of an length-$h$ outer code $\outcode$ whose task is to generate an output sequence with a prescribed empirical distribution. Following \cite{bocherer2015bandwidth}, we refer to such a device as the \ac{DM}. We restrict out attention to \acp{DM} based on \ac{CC} codes which admit low-complexity implementation via arithmetic coding \cite{Mac99,Schulte2016} . Let $\omega\in(0,1)$ denote the fractional Hamming weight of the $h$-bits vector $\vecv$, i.e., $\omega = \hw(\vecv)/h$. We have that $M = |\outcode|={h \choose \omega h}$. Hence, the rate of the outer code is $\dmR = h^{-1}\log_2 M = h^{-1}\log_2 {h \choose \omega h}$, which converges to $\Hb(\omega)$ for large $h$. The output of the \ac{DM} is then input to the encoder of an inner $(n,h)$ binary linear block code $\incode$. The inner code $\incode$ is defined by
\begin{equation}
\incode = \left\{ \vecc \big| \vecc \vH_2^\transposed = \vecv \vH_1^\transposed, \vecv \in \fieldtwo^h\right\}
\end{equation}
where $\vH_1$ is an $n \times h$ sparse binary matrix, and $\vH_2$ is an $n \times n$ sparse binary matrix. Note that, strictly speaking, $\incode$ may not be an \ac{LDPC} code, i.e., the code may not possess a sparse parity-check matrix. Nevertheless, $\incode$ can be seen as the code obtained by puncturing an $(n+h,h)$ \ac{LDPC} code with $n\times (h+n)$ parity-check matrix 
\begin{equation}
\vH = \left[\,\vH_1\, |\, \vH_2\,\right] \label{eq:LDPCH}
\end{equation}
where puncturing is applied to the first $h$ coordinates. We refer to the $(n+h,h)$ \ac{LDPC} code with parity-check matrix in the form \eqref{eq:LDPCH} as the (inner) \emph{mother code} $\inmcode$. 
The inner code rate is $\inR = h/n$, whereas the mother code rate is $\inmR = h / (n+h) = \inR / (1 + \inR)$. We assume $\vH_2$ to possess rank $n$, implying that $\vH_2$ is invertible. We denote by $\bm{G}=\vH_1 \vH_2^{-\transposed}$ the generator matrix of the inner code $\incode$. Note that if the inverse of $\vH_2$ is dense, then the generator matrix $\bm{G}$ is dense. It follows that the concatenation of the outer \ac{CC} code with the inner linear block code yields a marginal distribution of the codeword bits that is close to uniform, as required by the capacity-achieving input distribution of the \ac{biAWGN} channel \cite{bocherer2015bandwidth}.
The overall code obtained by the concatenation of the outer \ac{CC} code and the inner binary linear block code is denoted by $\code$, and its rate is $R = \dmR \inR \approx \Hb(\omega) \inR$. We remark that all rates $0<R<\inR$ can be achieved simply by fixing the \ac{DM} parameter, hence its rate, without performing any modification (e.g., puncturing/shortening) to the inner code.

We consider \ac{MN} codes based on a protograph mother \ac{LDPC} code $\inmcode$. In particular, the mother code parity-check matrix \eqref{eq:LDPCH} is obtained by lifting a protograph whose $n_0 \times (h_0 + n_0)$ base matrix takes the form
$\vB = \left[\,\vB_1\, |\, \vB_2\,\right]$
where $\vB_1$ has dimensions $n_0 \times h_0$ and $\vB_2$ has dimensions $n_0 \times n_0$, where $n_0=n/\liftfac$ and $h_0=h/\liftfac$ are positive integers and where $\liftfac$ is the integer lifting factor. Note that all \acp{VN} of type $i$ with $i=1,\ldots,h_0$ are punctured. 
An \ac{MN} code is fully defined by the parity-check matrix of the inner mother code $\inmcode$ and by the \ac{DM} parameter $\omega$. We refer to a \ac{MN} code \emph{family} as the set of \ac{MN} codes with fixed mother code, obtained for all  $\omega \in (0,1)$ that yield an integer $\omega h$. Similarly, we refer to a (protograph) \ac{MN} code ensemble  $ \ensP{\pgraph}{\omega} $ as the set of \ac{MN} codes whose inner mother code bipartite graph $\bgraph$ is obtained by lifting $\pgraph$, and where the \ac{DM} paramerer is $\omega$. The \ac{MN} code ensemble family $\ensfP{\pgraph}$ is the set of ensembles  $\left\{\ensP{\pgraph}{\omega}\right\}_{\omega \in (0,1)}$.

\begin{figure*}
		\centering
		{
		\includegraphics[width=16cm]{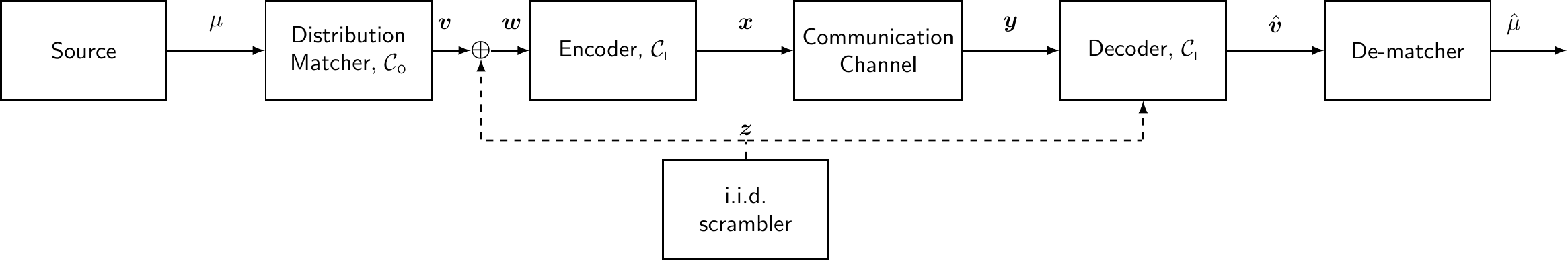}
        }
        \vspace{-1mm}
  \caption{Modification of the system model of Figure \ref{fig:model}, where an i.i.d. scrambler is introduced.}
\label{fig:equiv1}
\end{figure*} 		

\begin{figure}
		\centering
		{
		\includegraphics[width=0.9\columnwidth]{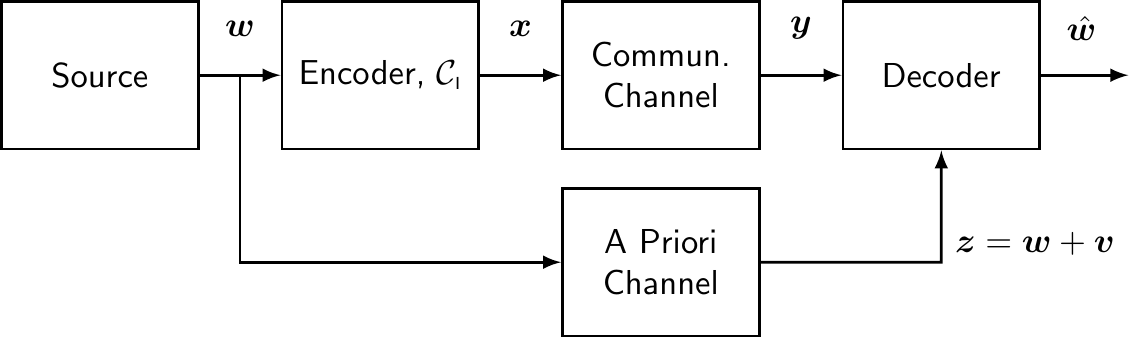}
		}
  \vspace{-1mm}
	\caption{Equivalent parallel channel model.}
	\label{fig:equiv2}
 \vspace{-3mm}
\end{figure}  

\subsection{Belief Propagation Decoding}\label{sec:MNcodes:BPdecoding}
At the channel input, each codeword $\vecc$ is mapped onto $\{-1,+1\}^n$ via binary antipodal modulation, i.e., $x_i = 1-2c_i$ for $i=1,\ldots,n$. 
With a slight abuse of notation, we will refer to the modulated codeword $\vecx$ as the codeword. We will also use $\incode$ and $\code$ to denote the modulated codebook of the inner code and of the overall code, respectively.
We assume next that decoding is performed via \ac{BP} applied to the bipartite graph of the mother \ac{LDPC} code.
We initialize the \ac{BP} decoder as follows. Let us denote by $L_i$ the $L$-value at the input of the $i$th \ac{VN}. Moreover, define $\Delta = \ln [(1-\omega)/\omega]$. We set $L_i = \Delta$ is $i\leq h$, whereas we set $L_i = 2 y_{i-h}/\sigma^2$ if $i>h$. In other words, for the punctured \acp{VN} we provide prior information obtained from the marginal distribution of the \ac{CC} code codeword $\vecv$, while for \acp{VN} associated with the codeword bits that are transmitted through the \ac{biAWGN} channel, we input the corresponding channel \acp{LLR}. The \ac{BP} decoder output $\hat{\vecv}$ in then processed by the de-matcher \cite{Schulte2016}, producing the estimate $\hat{\mu}$ of the transmitted message (see Figure \ref{fig:model}). It is interesting to note that this decoder (proposed already in \cite{Mac99}) employs the same layered decoding architecture \cite{bocherer2017achievable} adopted by \ac{PAS} \cite{bocherer2015bandwidth}: the \ac{BP} decoder does not have any information on the outer \ac{CC} constraints, and it exploits only the knowledge of the marginal distribution of the bits in $\vecv$.

\subsection{More on Decoding Metrics}\label{sec:MNcodes:decoding_metrics}

We introduce next two decoding approaches which, albeit impractical, will be useful in the analysis of \ac{MN} codes. We first consider the \ac{ML}  decoder
\begin{equation}
\xML = \argmax_{\vecx\in \code} p(\vecy|\vecx) \label{eq:ML1}
\end{equation}
where $p(\vecy|\vecx)$ is the probability density of the \ac{biAWGN} channel output $\vecy$ conditioned on the input $\vecx$. Note that the \ac{ML} criterion \eqref{eq:ML1} can be rephrased as
\begin{equation}
\xML = \argmax_{\vecx\in \incode} p(\vecy|\vecx(\vecv))P(\vecv). \label{eq:ML2}
\end{equation}
In \eqref{eq:ML2}, the dependency of $\vecx$ (inner encoder output) on $\vecv$ (inner encoder input) is emphasized. Note in particular that the search is over the inner code $\incode$, hence, over an enlarged set compared to \eqref{eq:ML1}: the overall code structure is conveyed by the prior $P(\vecv)$ taking value $1/|\outcode|$ if $\vecv \in \outcode$, and taking value  zero otherwise. Following the spirit of the \ac{BP} decoder, which employs the marginal distribution of the bits composing $\vecv$ to bias the decoder operating over the bipartite graph of the mother code, we consider also a decoder that outputs
\begin{equation}
\xMM = \argmax_{\vecx\in \incode} p(\vecy|\vecx(\vecv))Q(\vecv). \label{eq:MM1}
\end{equation}
where $Q(\vecv)=\prod_{i=1}^h P(v_i)$ denotes the product of the marginal distributions of the bits $v_1, v_2,\ldots,v_h$. 
The decoding metric adopted in \eqref{eq:MM1} is clearly suboptimal compared with the one of \eqref{eq:ML2}. In fact, the term $Q(\vecv)$ acts as a mismatched prior, yielding a nonzero probability also for $\vecv \notin \outcode$. We refer to the decoding metric $p(\vecy|\vecx(\vecv))Q(\vecv)$ as the \emph{mismatched  metric}.

\subsection{Equivalent Parallel Channel Model}\label{sec:MNcodes:equivmodels}
The analysis of the scheme described in the previous sections presents some challenges. For example, being the code $\code$ nonlinear, the block error probability under \eqref{eq:ML1}--\eqref{eq:ML2} will depend upon the transmitted codeword. This hinders the use of a reference codeword to compute bounds on the block error probability. A similar issue arises when attempting a \ac{DE} evolution analysis under \ac{BP} decoding, where typically the allzero codeword is used as reference. The issue can be circumvented by resorting to alternative communication models that can be proved to be equivalent to the one depicted in Figure \ref{fig:model} \cite{wyner1974recent,fresia2010joint}. Consider first the scheme depicted in Figure \ref{fig:equiv1}, where a scrambling block is introduced. The block generates a sequence $\vecz=(z_1,z_2,\ldots,z_h)$ where each element is picked independently and uniformly at random in $\{0,1\}$. The sequence $\vecz$ is then added (in $\fieldtwo$) to $\vecv$ prior to encoding with $\incode$. The same sequence is made available to the decoder. Considering either \eqref{eq:ML1} or \eqref{eq:ML2}, and owing to the symmetry of the \ac{biAWGN} channel, we observe that the presence of the scrambler is irrelevant to the analysis of the error probability, since the addition of $\vecz$ at the transmitter side can be compensated at the decoder by computing first $\vecb=\vecz \bm{G}$, and then flipping the sign of the observations $y_i$ for all $i\in \mathrm{supp}(\vecb)$. The model of Figure \ref{fig:equiv1} admits an equivalent model, provided in Figure \ref{fig:equiv2}: here, an \ac{i.i.d.} uniform binary source generates an $h$-bits vector $\vecw$, which is encoded via $\incode$ yielding a codeword $\vecx$ that is transmitted through the \ac{biAWGN} channel. The decoder obtains also an observation of $\vecw$ via a so-called \emph{a priori channel}. The a priori channel adds (in $\fieldtwo$) a weight-$\omega h$ binary vector $\vecv$ to $\vecw$, were $\vecv$ is picked uniformly at random in $\outcode$, resulting in the observation $\vecz$. Upon observing $\vecy$ and $\vecz$, the decoder produces a decision on $\vecw$, or, equivalently, a decision on $\vecv$ since $\vecw=\vecv+\vecz$. \ac{ML} decoding will produce
\begin{equation}
\xML = \argmax_{\vecx\in \incode} p(\vecy|\vecx(\vecw))P(\vecz|\vecw) \label{eq:ML3}
\end{equation}
where $P(\vecz|\vecw)=1/|\outcode|$ if $(\vecz-\vecw) \in \outcode$, and $P(\vecz|\vecw)=0$ otherwise. The decoding problem can be immediately recognized to be equivalent to the one in \eqref{eq:ML2}. The decoder may also resort to a mismatched model for the a priori channel, treating it as a \ac{BSC} with crossover probability $\omega$ and resulting in 
\begin{equation}
\xMM = \argmax_{\vecx\in \incode} p(\vecy|\vecx(\vecw))Q(\vecz|\vecw) \label{eq:ML4}
\end{equation}
where $Q(\vecz|\vecw)=\prod_{i=1}^h P(z_i|w_i)$, i.e., the solution of \eqref{eq:ML4} is equivalent to the solution of \eqref{eq:MM1}.
We refer to the model of Figure \ref{fig:equiv2} as the \ac{EPC} \emph{model}. The convenience of the \ac{EPC} model stems from the fact that, owing to the symmetry of the communication and of the a priori channels, and to the linearity of the code $\incode$, the error probability is independent on $\vecw$: we can analyze the error probability of the scheme of Figure \ref{fig:equiv2} by fixing as reference the allzero codeword. The resulting analysis will characterize exactly the performance of the original scheme of Figure \ref{fig:model}.
\section{Density Evolution Analysis}\label{sec:DE}

Protograph-based \ac{MN} code ensembles can be analyzed via \ac{DE} by resorting to the \ac{EPC} model introduced in Section \ref{sec:MNcodes:equivmodels}. The analysis shares several commonalities with the \ac{DE} of \ac{LDPC} code ensembles designed for joint source and channel coding \cite{fresia2010joint}. We performed the analysis both via quantized \ac{DE} \cite{RU01a}, as well as by means of the Gaussian approximation in the form of \ac{PEXIT} analysis \cite{ten01,Liva2007:EXIT_GC}. In either case, with reference to the \ac{EPC} model, we replaced the a priori channel (that introduces a constant number of errors $\omega h$ in $\vecz$) with a \ac{BSC} with crossover probability $\omega$. The choice is justified by observing that, as $h$ (and $n$) grows large, the fraction of errors introduced by the \ac{BSC} concentrates around $\omega$. We describe next how to perform the \ac{PEXIT} analysis, following the steps in \cite{Liva2007:EXIT_GC}. 
Denote by $\cbsc(\omega)$ the capacity of a \ac{BSC} with crossover probability $\omega$, and by $\cawgn(E_s/N_0)$ the capacity of a \ac{biAWGN} channel with \ac{SNR} $E_s/N_0$. The \ac{PEXIT} analysis  is initialized by setting the \ac{MI} at the input of type-$i$ \acp{VN}, $i=1,\ldots,h_0$, to $\cbsc(\omega)$, whereas for $i=h_0+1,\ldots, h_0+n_0$ the \ac{MI} at the input of type-$i$ \acp{VN} is initialized to $\cawgn(E_s/N_0)$. The analysis in then carried out via the recursions provided in \cite{Liva2007:EXIT_GC}, and it allows to determine (for fixed $\omega$) the iterative decoding threshold over the \ac{biAWGN} channel, that is the minimum $E_s/N_0$ for which the \ac{MI} values tracked by the \ac{PEXIT} analysis converge to $1$. We denote the threshold value  as $\thr(R)$, where we emphasize the dependency on $R$ (and, hence, on $\omega$).
Due to the faster computations entailed by the \ac{PEXIT} analysis with respect to quantized \ac{DE}, \ac{PEXIT} analysis will be used for protograph optimization, while quantized \ac{DE} will be used to verify the accuracy of the thresholds computed for the designed protographs.

The threshold $\thr$ can be used as target measure to be minimized in the design of protographs. In particular, suppose we are interested in finding an inner mother code protograph that allows to operate close to capacity over a range $\left[\rmin,\rmax\right]$ of rates $R$, i.e., over a range $\left[\omin,\omax\right]$ of values of $\omega$.
Fix the protograph parameters $h_0, n_0$. By doing so, the inner code rate $\inR$ is fixed.  A set of target rates $\rset \subset \left[\rmin,\rmax\right]$   is selected. For each target rate $R\in\rset$ we can derive the rate of the \ac{DM} as $\dmR = R/\inR$, out of which the \ac{DM} parameter $\omega$ is obtained. Given a protograph $\pgraph$ we define the \ac{WCL} 
\begin{equation}
\Delta(\pgraph) := \max_{R \in \rset} \left[\thr(R) - \cawgn^{-1}(R)\right]\label{eq:WCL}
\end{equation}
that is the maximum gap between the protograph iterative decoding threshold and the \ac{biAWGN} Shannon limit  for the rates in $\rset$.
A search for the protograph with parameters $h_0, n_0$ that minimizes the \ac{WCL} in \eqref{eq:WCL} can be carried out, for example, via differential evolution \cite{shokrollahi2005design}. We provide next some examples of application to the design of protograph-based \ac{MN} code ensemble families addressing different rate regimes. In all examples, the search space was limited by setting the maximum number of parallel edges between protograph \ac{VN}/\ac{CN} pairs (i.e., value of the base matrix elements) to $3$.

\begin{example}\label{ex:12}
    Consider a code rate range $[0.1, 0.5]$. An \ac{MN} code family addressing this range can be derived from an inner $\inR=1/2$ code, i.e., the mother code has rate $\inmR=1/3$. We search for protographs with $6$ \acp{VN} and $4$ \acp{CN}, minimizing the \ac{WCL} over $\rset=\{0.1, 0.3, 0.5\}$. We obtain the base matrix
\begin{align}
    \vB_{1/2} &= \left(
    \begin{array}{cc|cccc}
     1 &  0 & 1 & 1 & 0 & 0\\
     0 &  1 & 0 & 3 & 0 & 1\\
     2 & 0 & 1 & 1 & 1 & 0\\
     1 & 2 & 1 & 2 & 0 & 0
    \end{array}
    \right).
\end{align}
where the first two columns are associated with punctured \acp{VN}.
The decoding thresholds for the code ensemble defined by $\vB_{1/2}$ are provided in Table~\ref{tab:DEvsEXIT} for various rates. The values are computed with both quantized \ac{DE} and \ac{PEXIT} analysis. The results confirm the accuracy of the latter, with thresholds that are within $0.05$ dB from the quantized \ac{DE} ones for rates in the ranging from $0.3$ to $0.5$. The accuracy reduces for the lowest rates, however still yielding acceptable estimates.  
\end{example}

\bgroup
\def\arraystretch{1.15}
\begin{table}[t]
	\caption{Thresholds (\textnormal{d}B) computed by quantized DE and by PEXIT for different overall code rate $R$}
	\centering
    \vspace{-0mm}
\begin{tabular}{c c c c c c}
			\hline\hline
            {$R$} & $0.5$ & $0.4$ & $0.3$ & $0.2$ & $0.1$ \\ \hline
            $\thr$ dB (quant. \ac{DE}) & $-2.01$  & $-3.38$  & $-5.02$  & $-7.00$  & $-10.32$  \\
            $\thr$ dB ({PEXIT}) & $-2.06$  & $-3.42$  & $-5.05$  & $-7.14$  & $-10.49$  \\ \hline\hline
	\end{tabular}
	\label{tab:DEvsEXIT}
 \vspace{-0mm}
\end{table}
\egroup

\begin{example}\label{ex:23}
    Consider a code rate range $[0.1, 0.666]$. An \ac{MN} code family addressing this range can be derived from an inner $\inR=2/3$ code. Accordingly, fix the base matrix to be a $3 \times 5$ matrix. We search for protographs minimizing the \ac{WCL} over $\rset=\{0.1, 0.3, 0.666\}$. We obtain the base matrix
\begin{align}
    \vB_{2/3} &= \left(
    \begin{array}{cc|ccc}
      1 &  0 & 0 & 3 & 1\\
     1 &  1 & 0 & 3 & 0\\
     1 & 2 & 2 & 1 & 0
    \end{array}
    \right).
\end{align}
\end{example}

\begin{figure}
		\centering
		{
		\includegraphics{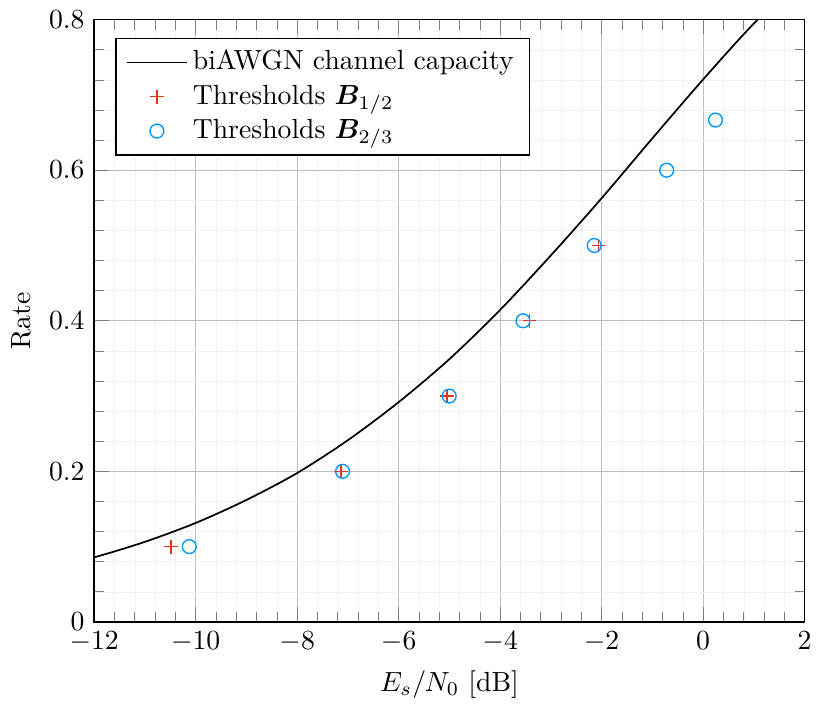}
        }
  \caption{Iterative decoding thresholds computed for the \ac{MN} code ensembles defined by the base matrices described in Example \ref{ex:12} and in Example \ref{ex:23}.}
\label{fig:thresholds}
\end{figure}

Remarkably, the iterative decoding thresholds for both ensembles  of Examples \ref{ex:12} and \ref{ex:23} (displayed in Figure \ref{fig:thresholds}) are within $1$ dB from the Shannon limit over a wide range of rates.
\section{Finite Length Performance Analysis}\label{sec:ErrorFloor}

In this section, we provide numerical results on finite-length constructions of protograph \ac{MN} codes. We complement Monte Carlo simulation results with an error floor analysis based on the \ac{UB} on the block error probability. In particular, we analyze mismatched decoding as defined in \eqref{eq:ML4}. This choice follows the observation that the \ac{BP} decoder does not exploit the joint \ac{p.m.f.} $P(\vecv)$, but rather the marginal distribution $P(v_i)$ for all $i=1,\ldots,h$. By resorting to the \ac{EPC} setting, the derivation of bounds on the error probability under \eqref{eq:ML3} reduces to the analysis of the error probability under \eqref{eq:ML4}. We first derive the pairwise error probability
\begin{equation}
\PEP(\vecx') = \mathsf{P} \big[ \,p(\vecY|\vecx)Q(\vecZ|\vecw)\leq p(\vecY|\vecx')Q(\vecZ|\vecw')\,\big]. \label{eq:PEP1}
\end{equation}
In \eqref{eq:PEP1}, the codeword transmitted over the communication channel is $\vecx$, and it is the result of the encoding of $\vecw$, where the vector $\vecw$ is transmitted over the a priori channel. The competing codeword is $\vecx'$,  and it is the result of the encoding of $\vecw'$. Note that in \eqref{eq:PEP1} ties are broken in favor of the competing codeword. Owing to the symmetry of the communication and a priori channels, and to the linearity of $\incode$, we assume without loss of generality that $\vecw = (0, 0, \ldots, 0)$ and hence $\vecx=(+1, +1, \ldots, +1)$. Conditioned on $\bm{X}=\vecx$ and $\bm{W}=\vecw$, $\vecZ$ is uniformly distributed over the set of $h$-bit sequences with Hamming weight $\omega h$, whereas $Y_1, Y_2, \ldots, Y_n$ are \ac{i.i.d.} $\sim \mathcal{N}(+1,\sigma^2)$.
We can rewrite \eqref{eq:PEP1} as
\begin{align}
\PEP(\vecx') 
             &= \prob{\sum_{i=1}^n \ln  \frac{p(Y_i|x_i)}{p(Y_i|x_i')} \leq \sum_{i=1}^h \ln  \frac{Q(Z_i|w_i')}{Q(Z_i|w_i)}}\\
             &= \prob{\textstyle \sum_{i\in \mathcal{D}(\vecx')}  L_i \leq - \sum_{i\in \mathcal{D}(\vecw')} T_i}.
\label{eq:PEP2}
\end{align}
In \eqref{eq:PEP2}, $\mathcal{D}(\vecx')=\{i | x_i' \neq x_i\}$, $\mathcal{D}(\vecw')=\{i | w_i' \neq w_i\}$, whereas $L_i := \ln [ p(Y_i|+1)/p(Y_i|-1)]$ and $T_i := \ln [ p(Z_i|0)/p(Z_i|1)]$.  Denote by $L = \sum_{i\in \mathcal{D}(\vecx')}  L_i$ and by  $T = \sum_{i\in \mathcal{D}(\vecw')} T_i$. Moreover, let
$\delta_1 = \hd(\vecw,\vecw')$ and $\delta_2 = \hd(\vecx,\vecx')$. 
Conditioned on $\bm{X}=\vecx$ and $\bm{W}=\vecw$, we have that $L\sim\mathcal{N}(2\delta_2/\sigma^2,4\delta_2/\sigma^2)$. Recalling $\Delta = \ln [(1-\omega)/\omega]$, we have that $T_i = \Delta$ if $Z_i=0$, whereas $T_i = -\Delta$ if $Z_i=1$, i.e.,
\[
T = (\delta_1 - E)\Delta - E\Delta = \delta_1 -2E\Delta
\]
where $E$ follows an hypergeometric distribution with parameters $(h,\omega h, \delta_1)$.
After a few simple manipulations, we obtain 
\begin{equation}
\PEP(\vecx') = \expect{Q\left(\frac{2\delta_2/\sigma^2 + \delta_1 -2E\Delta}{2\sqrt{\delta_2}/\sigma}\right)}\label{eq:PEPfin}
\end{equation}
where $Q(x)$ is the well-known Gaussian $Q$-function.
By observing that $\PEP(\vecx')$ depends on $\vecx'$ only through its Hamming distance from $\vecx$, and on the Hamming distance between the corresponding information sequence $\vecw'$ and $\vecw$, we can upper bound the block error probability under \eqref{eq:ML4} as 
\begin{align}
P_B &\leq 
 \sum_{\delta_1 = 1}^{h} \sum_{\delta_2 = 1}^{n} \iowe_{\delta_1,\delta_2}  \expect{Q\left(\frac{2\delta_2/\sigma^2 + \delta_1 -2E\Delta}{2\sqrt{\delta_2}/\sigma}\right)} \label{eq:UB}
\end{align}
where $\iowe_{\delta_1,\delta_2}$ is the input-output weight enumerator of $\incode$. At large \acp{SNR}, \eqref{eq:UB} can be approximated by truncating the summation to the dominant term $\iowe_{\delta^\star_1,\delta^\star_2}$, yielding
\begin{align}
P_B \approx \iowe_{\delta^\star_1,\delta^\star_2}  \expect{Q\left(\frac{2\delta^\star_2/\sigma^2 + \delta^\star_1 -2E\Delta}{2\sqrt{\delta^\star_2}/\sigma}\right)}. \label{eq:TUB}
\end{align}
A possible strategy to identify the dominant term $\iowe_{\delta^\star_1,\delta^\star_2}$ can be to enumerate low-weight codewords of $\incode$ (e.g., via the algorithm proposed in \cite{Hu2004a}). Once a sufficient number of low-weight codewords is collected, their contribution to the error probability can be measured via \eqref{eq:PEPfin}.

\begin{figure}[t]
\centering
\includegraphics{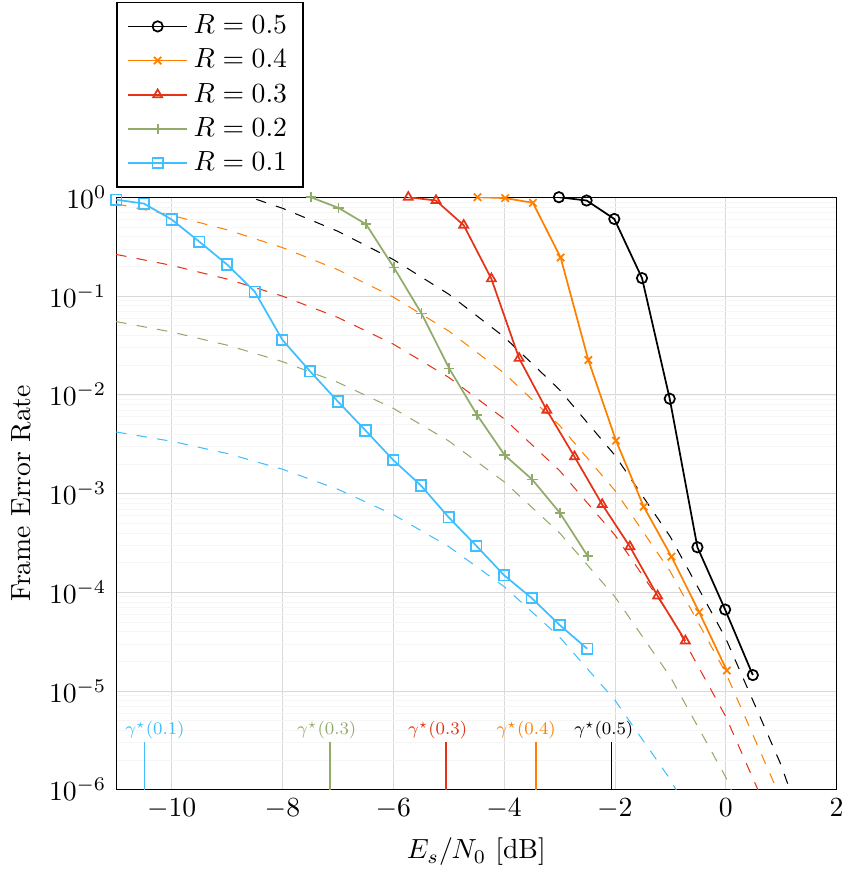}
\caption{\ac{FER} vs. $E_s/N_0$ dB for length-$1200$ protograph \ac{MN} code with inner mother code base matrix $\vB_{1/2}$ and code rate range $[0.1, 0.5]$. The \ac{TUB} of the block error probability for each rate is provided (dashed lines) as well as the corresponding iterative decoding thresholds.}
\label{fig:FER_UB}
\end{figure}

Figure \ref{fig:FER_UB} reports simulation results for a length-$1200$ \ac{MN} code family, for different rates $R\in[0.1,0.5]$. The base matrix of $\inmcode$ is $\bm{B}_{1/2}$. The protograph was lifted through a circulant version of the \ac{PEG} algorithm \cite{HEA05}. The \acp{TUB} provide an excellent prediction of the \ac{FER} at large \acp{SNR}. Interestingly, the \acp{TUB} indicate at large \ac{SNR} a diminishing return in coding gain when the rate of the outer \ac{CC} is reduced, whereas the coding gains at moderate \ac{FER} are more sizeable.

\section*{Acknowledgment}
The authors would like to thank Prof. Gerhard Kramer (Technical University of Munich) for his constructive comments and suggestions.




\balance
\end{document}